\documentclass[superscriptaddress,
amsmath,amssymb,
aps,prb,reprint,
]{revtex4-2}

\usepackage{graphicx}
\usepackage{bm}
\usepackage{amsmath}
\usepackage{miller}

\usepackage{siunitx}
\DeclareSIUnit{\oersted}{Oe}

\begin{document}
\title{Quantum-well tunneling anisotropic magnetoresistance above room temperature}
\date{\today}

    \author{Muftah Al-Mahdawi}
    \email{mahdawi@mlab.apph.tohoku.ac.jp}
    \thanks{These authors have equal contributions.}
    \affiliation{National Institute for Materials Science, Tsukuba 305-0047, Japan}
    \affiliation{Center for Science and Innovation in Spintronics (Core Research Cluster), Tohoku University, Sendai 980-8577, Japan}
    \affiliation{Center for Spintronics Research Network, Tohoku University, Sendai 980-8577, Japan}
    \author{Qingyi Xiang}
    \thanks{These authors have equal contributions.}
    \affiliation{National Institute for Materials Science, Tsukuba 305-0047, Japan}
    \author{Yoshio Miura}
    \email{miura.yoshio@nims.go.jp}
    \thanks{These authors have equal contributions.}
    \affiliation{National Institute for Materials Science, Tsukuba 305-0047, Japan}
    \affiliation{Center for Spintronics Research Network, Osaka University, Toyonaka 560-8531, Japan}
    \author{Mohamed Belmoubarik}
    \affiliation{National Institute for Materials Science, Tsukuba 305-0047, Japan}
    \author{Keisuke Masuda}
    \affiliation{National Institute for Materials Science, Tsukuba 305-0047, Japan}
    \author{Shinya Kasai}
    \affiliation{National Institute for Materials Science, Tsukuba 305-0047, Japan}
    \author{Hiroaki Sukegawa}
    \affiliation{National Institute for Materials Science, Tsukuba 305-0047, Japan}
    \author{Seiji Mitani}
    \email{mitani.seiji@nims.go.jp}
    \affiliation{National Institute for Materials Science, Tsukuba 305-0047, Japan}
    \affiliation{Graduate School of Pure and Applied Sciences, University of Tsukuba, Tsukuba 305-0047, Japan}

\begin{abstract}
    Quantum-well (QW) devices have been extensively investigated in semiconductor structures. More recently, spin-polarized QWs were integrated into magnetic tunnel junctions (MTJs). 
    In this work, we demonstrate the spin-based control of the quantized states in iron $3d$-band QWs, as observed in experiments and theoretical calculations. We find that the magnetization rotation in the Fe QWs significantly shifts the QW quantization levels, which modulate the resonant-tunneling current in MTJs, resulting in a tunneling anisotropic magnetoresistance (TAMR) effect of QWs. This QW-TAMR effect is sizable compared to other types of TAMR effect, and it is present above the room-temperature.
    In a QW MTJ of Cr/Fe/MgAl$_2$O$_4$/top electrode, where the QW is formed by a mismatch between Cr and Fe in the $d$ band with $\Delta_1$ symmetry, a QW-TAMR ratio of up to \SI{5.4}{\percent} was observed at \SI{5}{\kelvin}, which persisted to \SI{1.2}{\percent} even at \SI{380}{\kelvin}. 
    The magnetic control of QW transport can open new applications for spin-coupled optoelectronic devices, ultra-thin sensors, and memories. 
\end{abstract}

\maketitle

    The quantum well (QW) devices have found a wide adoption in semiconductor technologies, such as the QW-based lasers and high-mobility transistors \cite{esaki_1970,nag_2000}. On the other hand, metallic QWs can incorporate the large exchange splitting on $d$ electrons in spin-polarized QW devices \cite{garrison_1993,smith_1994,kawakami_1999,paggel_1999,luh_2000}. In more recent technological advances, spin-polarized $d$-band QWs were demonstrated in magnetic tunnel junctions (MTJs), where the confinement potential is produced both by band-gap and band-symmetry mismatches \cite{nozaki_2005,niizeki_2008,tao_2015,xiang_2019}.
    
    The MTJs have been the prominent driver in spintronics research and applications. Especially, the Fe-alloys/MgO MTJs possess a giant tunneling magnetoresistance (TMR) effect \cite{yuasa_2004,parkin_2004}, due to the symmetry-selective filtering of $d$ states \cite{butler_2001}. Another different effect is the anisotropic change of tunneling resistance of MTJs with the relative angle between tunneling current and magnetization direction. This tunneling anisotropic magnetoresistance (TAMR) effect stems from the changes in the density of states near Fermi level, due to spin-orbit coupling (SOC) effects \cite{gould2004,bolotin2006,gao2007,moser_2007,liu_2008,park_2008,park_2011,lu_2012,wang_2013a}. TAMR effect has been in interest for understanding the important role of SOC at FM interfaces, and for applications requiring ultra-thin spintronic sensors and memories, because of the need of only a single FM electrode.
    TAMR effect has been reported for large-SOC materials, such as GaMnAs \cite{gould2004,muneta_2017}, Co/Pt \cite{park_2008}, and IrMn \cite{park_2011}. In single-crystal Fe/MgO MTJs, TAMR effect was observed \cite{lu_2012}, and it was related to the combined effect of SOC and the asymmetric crystal field at interface \cite{khan_2008,lu_2012}.
    
\begin{figure}[!ht]
        \includegraphics[width=0.45\textwidth]{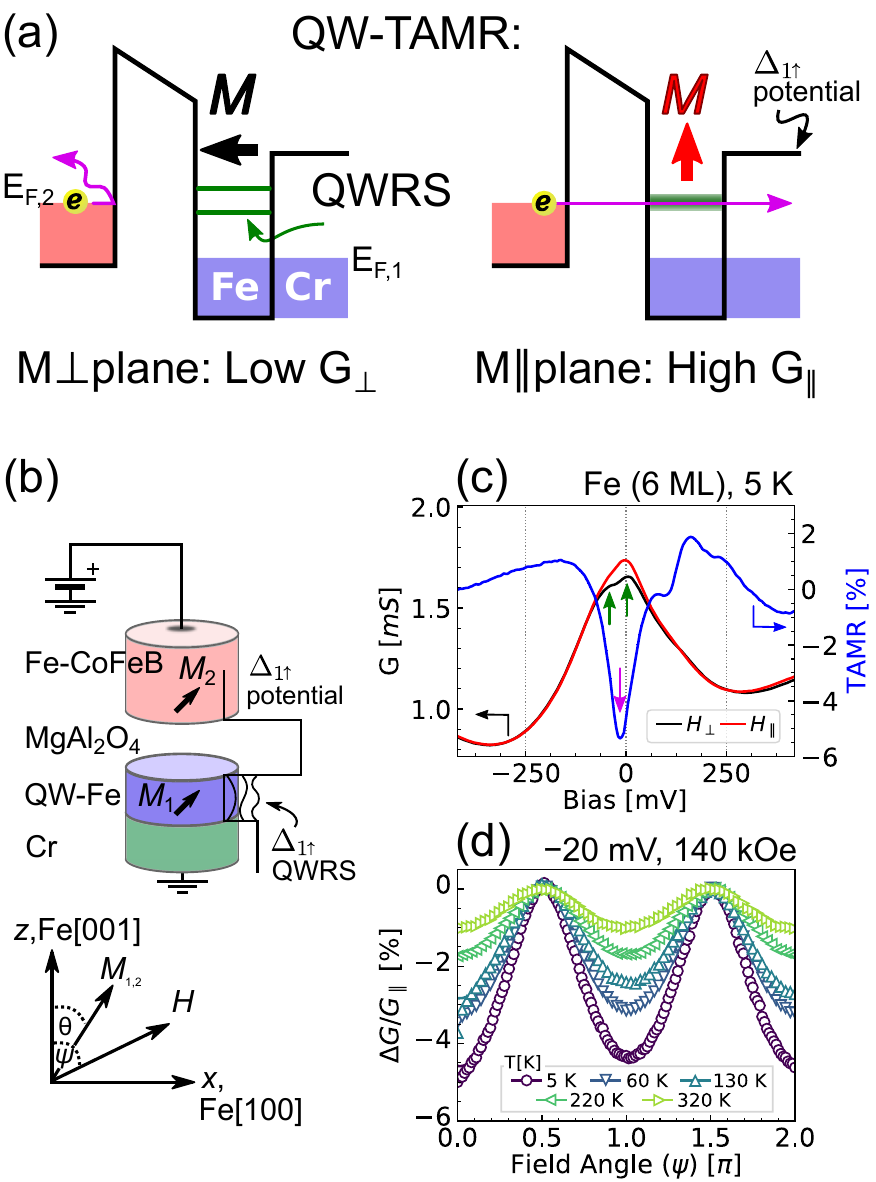}
    	\caption{The QW-TAMR effect. (a) In an Fe QW, the $M$ rotation shifts and broadens the sharp QW states, resulting in a large anisotropy of tunneling resistance (TAMR). (b) A schematic of the QW stack, and the definitions of coordinates. (c) The fine features of resonant conduction (green arrows) are anisotropic with respect to $M$ direction, resulting in a large QW-TAMR ratio (magenta arrow). (d) The QW-TAMR effect is twofold symmetric and persists above 300 K.
    	}
    	\label{fig:schem}
    \end{figure}

    In this work, we find a TAMR effect linked to the transport through QW resonant states (QWRSs) in a ferromagnetic QW structure. The narrow QWRSs couple to the magnetization direction through SOC, and the magnetization rotation controls the energy positions and broadening of QWRSs. Therefore, the magnetization anisotropically controls the resonant tunneling condition, and a sizable QW-TAMR effect is obtained [Fig.~\ref{fig:schem}(a)]. This QW-TAMR effect is qualitatively different from the much smaller TAMR effect in MTJs with bulk electrodes. Furthermore, the QW-TAMR effect is sustained above room temperature, compared to other TAMR effects, which were observed only at cryogenic temperatures.
    
    
    In the QW structure of Cr/Fe/oxide, the resonant tunneling is through QWRSs that form for the majority-spin electronic states of $\Delta_1$ symmetry ($\Delta_{1,\uparrow}$), due to a mismatch of $\Delta_1$ bands of Fe and Cr \cite{lu_2005,niizeki_2008}. In contrast to the rocksalt-type MgO, the spinel MgAl$_2$O$_4$ has a small lattice mismatch of $<\SI{1}{\percent}$ with Fe. A dislocation-free Fe/MgAl$_2$O$_4$ interface can be realized \cite{sukegawa_2010a,belmoubarik_2016}, which significantly enhances the phase coherence in QW structures \cite{tao_2015,xiang_2019}.
    We measure the QW-TAMR effect in epitaxial stacks of Cr/Fe/MgAl$_2$O$_4$, with the geometry and coordinates depicted in Fig.~\ref{fig:schem}(b). The detailed film stack is MgO-\hkl(001) substrate/MgO (5)/Cr (40)/Fe ($t_\text{Fe}$)/MgAl$_2$O$_4$ (2)/Fe (0.5)/Co$_{60}$Fe$_{20}$B$_{20}$(CoFeB) (5)/Ru (10) (thicknesses in nm), where $t_\text{Fe}$ = \num{0.70}, and \SI{0.84}{\nano\meter} correspond to 5 and 6 monolayers (ML), respectively. We used electron-beam evaporation and magnetron sputtering to prepare the epitaxial MTJ films. Post-annealing after each deposition step was done \emph{in situ} to improve crystallinity and flatness, except for the top Fe/CoFeB layer, which was left as deposited. Therefore, the top electrode provides an isotropic electrode for $\Delta_{1,\uparrow}$ electrons, due to the lack of crystallinity and the small TAMR in CoFeB \cite{gao2007}. Structural analysis showed a lattice-matched \hkl(001)-oriented epitaxial layer-by-layer growth and flat MgAl$_2$O$_4$ interfaces\cite{xiang_2017,xiang_2018}. The films were microfabricated into elliptical junctions with a cross section of \SI{5}{\micro\meter} by \SI{2.5}{\micro\meter}, where the major axis was along Fe \hkl[100] axis. Further details on the preparation and TMR properties are provided in Ref.~\cite{xiang_2019}. 
    As a comparison sample, we use a high-TMR non-QW sample epitaxially grown by sputtering with thick Fe electrodes \cite{belmoubarik_2016}, with the film stack of: MgO-\hkl(001) substrate/Cr (40)/Fe (100)/MgAl$_2$O$_4$ (2.17)/Fe (7)/IrMn (12)/Ru (10) (thicknesses in nm), which has a TMR ratio of \SI{401}{\percent} at \SI{5}{\kelvin}, and \SI{224}{\percent}  at \SI{300}{\kelvin}.
    
    We applied large saturating magnetic fields, and at each applied field angle $\psi$, we measured the $I$--$V$ curves at \SIrange{1}{5}{\milli\volt} steps with the four-wire method in a physical property measurement system, using sourcemeters and nanovoltmeters. Subsequently, we numerically calculated the differential tunnelling conductance $G = dI/dV$. The positive bias is defined as electrons tunnelling towards the top electrode. 
    
    At a Fe-QW thickness of \SI{6}{ML}, a QWRS forms near the Fermi level, as seen by the peak near zero bias in the $G$ spectra [Fig.~\ref{fig:schem}(c)]. The single $G$ resonance peak splits into two fine-structure peaks, when the magnetization $M$ is rotated by a saturating magnetic field ($H = \SI{140}{\kilo\oersted}$) from the inplane to out-of-plane direction [The green arrows in Fig.~\ref{fig:schem}(c)]. This split is the main origin of the QW-TAMR effect, that has a maximum magnitude in between the fine-structure peaks [The magenta arrow in Fig.~\ref{fig:schem}(c)]. The symmetry of the QW-TAMR effect is twofold with regard to out-of-plane field rotation angle [Fig.~\ref{fig:schem}(d)]. Furthermore, the QW-TAMR effect decreases gradually with increasing temperature, and is still present above the room temperature [Fig.~\ref{fig:schem}(d)]. We should note that there is a small fourfold component. This fourfold component is from the small misalignment of magnetization and field, and not intrinsic in origin \cite{chikazumi_2009_4fold}, as discussed in the Supplementary Materials \cite{NoteSupp}. Above the saturation field, there is no significant effect of $H$ magnitude (Sec.~S1 in \cite{NoteSupp}). Moreover, the QW-TAMR spectra are equivalent for $M$ rotations either in Fe \hkl(100) or \hkl(110) planes. The inplane \hkl(001) rotation has a zero QW-TAMR (Sec.~S2 in \cite{NoteSupp}). Therefore, the out-of-plane asymmetry is main component of the QWRS modulation.
    
    \begin{figure}[!ht]
        \includegraphics[width=0.45\textwidth]{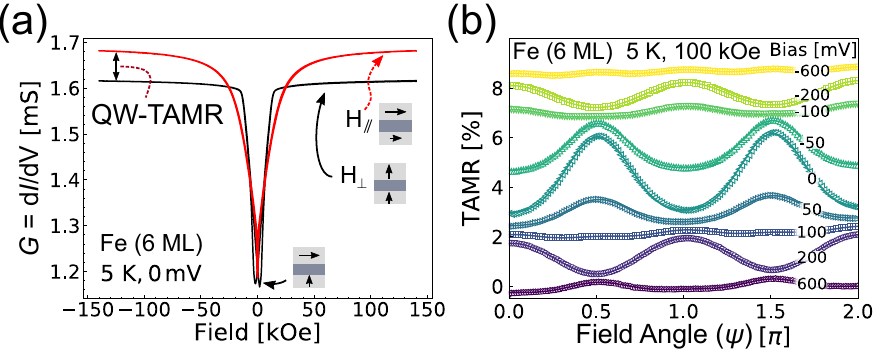}
    	\caption{Distinguishing QW-TAMR from TMR. (a) The $G$--$H$ curves show a large QW-TAMR effect between the inplane and perpendicular fields above saturation. (b) The angular dependence of TAMR show a strong dependence on bias. The solid lines are fitting curves to Eq.~\ref{eq:TAMR-angle}, and there is a vertical shift for clarity.
    	}
        \label{fig:RH_theta}
    \end{figure}
    
\begin{figure*}[!ht]
    	\includegraphics[width=1.0\textwidth]{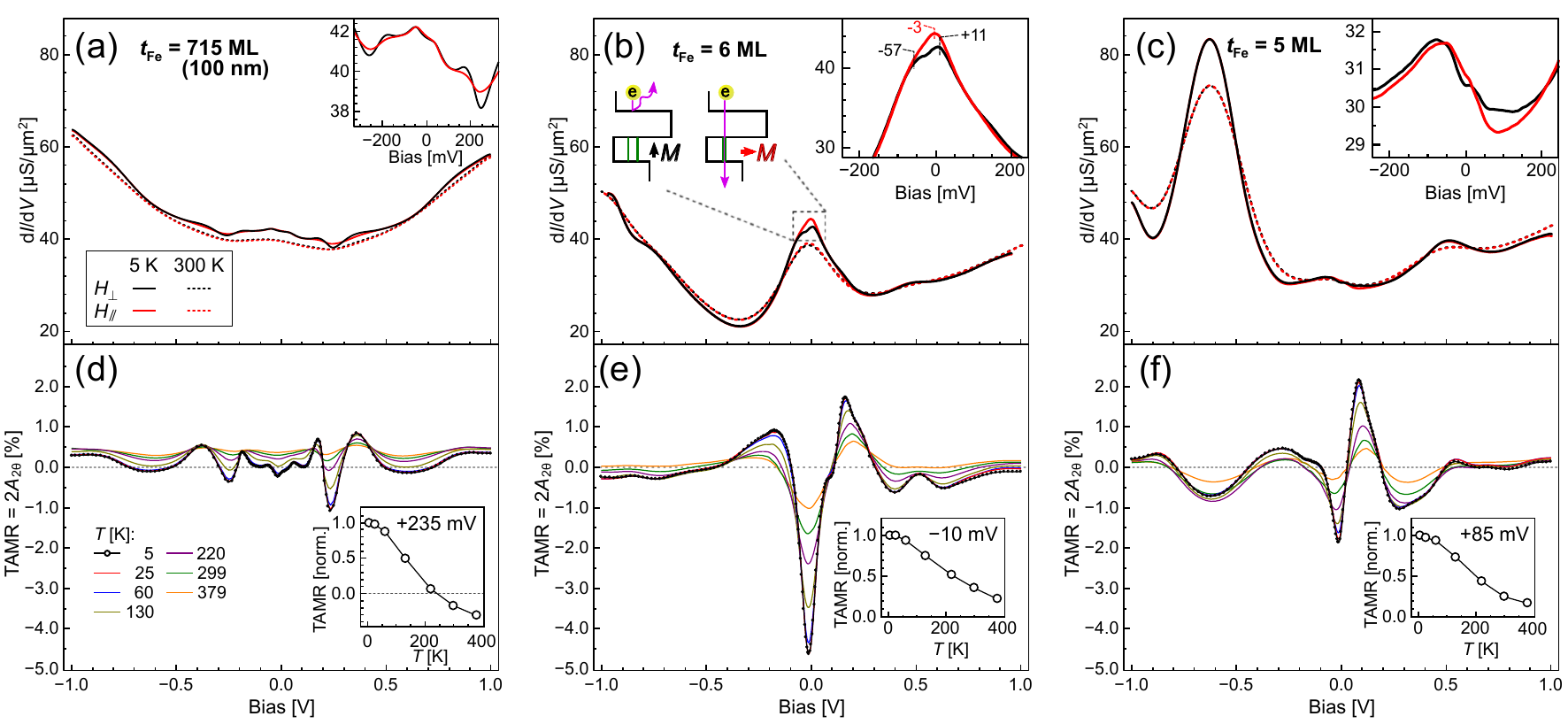}
    	\caption{The effect of QW formation on $G$ and TAMR spectra. (a--c) $G$ spectra at \SI{5}{\kelvin} (solid lines), and \SI{300}{\kelvin} (dashed lines). Black and red lines were obtained at saturating $H_\perp$ and $H_\parallel$, respectively. The insets are enlarged views near zero bias. The thickness of Fe $t_\text{Fe}$ is indicated on the figures. (d--f) The temperature dependence of TAMR spectra. The insets in (d--f) show the temperature dependence of largest TAMR peak normalized to its 5-K value.
    	}
    	\label{fig:TAMR}
    \end{figure*}

    In Figs.~\ref{fig:RH_theta} and \ref{fig:TAMR}, we show the distinction of QW-TAMR effect from the commonly observed TMR and TAMR effects. The $G$--$H$ curves in the inplane ($H_\parallel$) or out-of-plane ($H_\perp$) applied magnetic field directions are shown in Fig.~\ref{fig:RH_theta}(a). The low conductance at zero field is due to the TMR effect, and not of concern in this work. It is due to the non-parallel orthogonal magnetic configuration formed by the large perpendicular magnetic anisotropy at the bottom Fe/MgAl$_2$O$_4$ interface \cite{xiang_2018}.
    On the other hand, the top and bottom electrodes magnetization vectors $M_{1,2}$ are parallel at a high field above saturation ($\lvert H \rvert > \SI{50}{\kilo\oersted}$). The conductances $G$ in the $H_\parallel$ and $H_\perp$ field directions are not equal [Fig.~\ref{fig:RH_theta}(a)], resulting in an anisotropic tunneling conductance. This is what distinguishes TAMR, where conductance depends on the absolute angle $\theta$ of $M$, from the usual TMR effect, where conductance depends on relative angle between two magnetizations. Furthermore, the twofold nature of QW-TAMR effect is shown by the dominant twofold symmetry of the dependence of $G$ on field angle $\psi$, above satuarion field at $H = \SI{100}{\kilo\oersted}$  [Fig.~\ref{fig:RH_theta}(b)]. As the voltage bias is increased away from the QWRS near zero bias, the QW-TAMR shows a strong modulation in sign and magnitude, but the twofold symmetry is preserved. 
    We extract QW-TAMR spectra from fits to the $\psi$ dependence of $G$ at each bias, using the following definition of TAMR ratio:
    
    \begin{subequations}\label{eq:TAMR-angle}
    \begin{align}
    \mathrm{TAMR}(\psi) &= \frac{G(\psi)}{G(90^\circ)} - 1 \\
    					&= \mathrm{const.} + a_{2\psi} \cos 2\psi + a_{4\psi} \cos 4\psi \label{eq:G-psi} \\
    					&\equiv \mathrm{const.} + A_{2\theta} \cos 2\theta, \label{eq:G-theta}
    \end{align}
    \end{subequations}
    where $a_{2\psi}$ is the QW-TAMR apparent twofold component in $\psi$, $a_{4\psi}$ is the corresponding fourfold component. QW-TAMR shows only an intrinsic twofold symmetry $A_{2\theta}$ in the mangetization angle $\theta$, which is linked to $a_{2\psi}$ and $a_{4\psi}$ as follows (Sec.~S1 in \cite{NoteSupp}):
    
    \begin{equation}\label{eq:A2theta}
    A_{2\theta} \approx a_{2\psi} - 2 a_{4\psi}.
    \end{equation}
    
    Figure \ref{fig:TAMR} shows the distinction between the QW-TAMR and TAMR of bulk electrodes, the effect of QWRS position, and temperature dependence. We show the three cases of  a bulk-electrode sample ($t_\text{Fe} =$ \SI{715}{ML} = \SI{100}{\nano\meter}), and two QWs ($t_\text{Fe} =$ \numlist{6;5} \text{ML}). Figs.~\ref{fig:TAMR}(a--c) show the $G$--$V$ spectra in the $H_\parallel$ and $H_\perp$ field directions. Figs.~\ref{fig:TAMR}(d--f) show the $A_{2\theta}$--$V$ spectra extracted from Eq.~\ref{eq:A2theta}, including the temperature dependence.
    
    In contrast to the non-QW sample in Fig.~\ref{fig:TAMR}(a), the QW-Fe samples have the resonant tunneling peaks through QWRS near \SI{0}{\volt} for $t_\mathrm{Fe} =$ \SI{6}{ML}, and at \SI{-0.6}{\volt} for $t_\mathrm{Fe} =$ \SI{5}{ML} [Figs.~\ref{fig:TAMR} (b,c)]. 
    The QW resonance peaks in $G$--$V$ curves are sustained at room temperature and above [the dashed lines in Figs.~\ref{fig:TAMR} (b,c)]. The presence of QWRS at high temperatures is due to the phase coherence in Cr/Fe/MgAl$_2$O$_4$ QWs. The peak broadening of QWRS peaks by the carriers finite lifetime \cite{altfeder_1997, eisenstein_2007} is ranging \SIrange{0.23}{0.40}{\eV}. We estimate the maximum bound on majority carrier lifetime at \SI{2.8e-15}{\second} and the mean free path is \SI{3.3}{\nano\meter}, which are close to reported values on majority spin carriers in Fe \cite{enders_2001,niizeki_2008}. The ultra-thin QW produces a large separation between QWRS, on the order of \SIrange{0.5}{1.0}{\eV}, which is much larger than the QWRS broadening by the thermal energy or carrier lifetime.
     On the other hand, the non-QW sample shows a smoothing of $G$--$V$ curve features by increasing the temperature [Fig.~\ref{fig:TAMR}(a)].
    
    The QW formation and thickness cause a drastic change to TAMR spectra, as seen in Figs.~\ref{fig:TAMR} [d--f]. Also, relatively-large QW-TAMR magnitudes are observed at our maximum measurement temperature of \SI{380}{\kelvin} [Figs.~\ref{fig:TAMR} (e,f), and the insets therein]. In the non-QW sample, $G$--$V$ and TAMR--$V$ spectra have shallow valleys at \SI{\pm 0.25}{\volt} [Figs.~\ref{fig:TAMR}(a)]. The TAMR and $G$ spectra in thick-Fe/MgAl$_2$O$_4$ are qualitatively similar to MgO barrier \cite{lu_2012,tao_2018}, but more symmetric in bias direction and the TAMR ratio is comparatively larger due to better interfaces quality and lattice matching. When a QWRS forms close to the Fermi level at $t_\mathrm{Fe} =$ \SI{6}{ML}, we find a large QW-TAMR magnitude near zero bias [Fig.~\ref{fig:TAMR}(e)]. On the other hand, when $t_\mathrm{Fe} =$ \SI{5}{ML} and the QWRS level is at $0.6$-eV away from the Fermi level, QW-TAMR magnitude decreases at zero bias, while QW-TAMR has a small magnitude at the QWRS level of \SI{- 0.6}{\volt}. The quantization levels form either close to or far from zero bias depending on the parity of the number of Fe MLs, due to the Fe electronic structure \cite{lu_2005,niizeki_2008,xiang_2019}. The QW-TAMR magnitude correspondingly shows oscillations in $t_\mathrm{Fe}$ with a 2-ML period (Sec.~S3 in \cite{NoteSupp}). 
    
     The origin of QW-TAMR in QW samples can be considered as follows. The $M$ direction modifies the well potential via SOC, and causes a shift or a modification of the majority-to-majority spin conduction channel in the QW transport. As an example, we see a main single-peak feature in the 6-ML sample at \SI{-3}{\milli\volt} in the inplane magnetization direction, whereas a split spectrum is formed in the out-of-plane magnetization direction, with two fine-structure peaks at \SI{+11}{\milli\volt} and \SI{-57}{\milli\volt} [inset of Fig.~\ref{fig:TAMR}(b)]. Hence, a large anisotropy in $G$ is observed. Similarly at $t_\mathrm{Fe} = 5$ ML, there is an enhancement of TAMR at QWRS of \SI{-600}{\milli\volt}. Other shifts in the fine peaks at $t_\mathrm{Fe} = 5$ ML are also found [inset of Fig.~\ref{fig:TAMR}(c)]. 
     In contrast, the TAMR effect in thick-Fe/(MgO,MgAl$_2$O$_4$)/thick-Fe is from the relatively-small spin-flip transport channel dominated by the interfacial resonant states (IRSs) at an Fe surface \cite{butler_2001, lu_2012}. The IRSs are shifted when magnetization is rotated, causing a change in spin-flip conductances \cite{chantis2007,khan_2008,lu_2012}. This origin of thick-Fe TAMR is seen as a smoothing of $G$ spectra in the inset of Fig.~\ref{fig:TAMR}(a) upon the rotation of $M$. 
    To sum up, the modulation of the dominant $\Delta_1$ spin-conserved transport channel in an Fe-QW MTJ gives a larger QW-TAMR effect than the modulation of the much smaller IRS spin-flip transport channel of thick-electrode TAMR.

    \begin{figure}[!ht]
    	\includegraphics[width=0.45\textwidth]{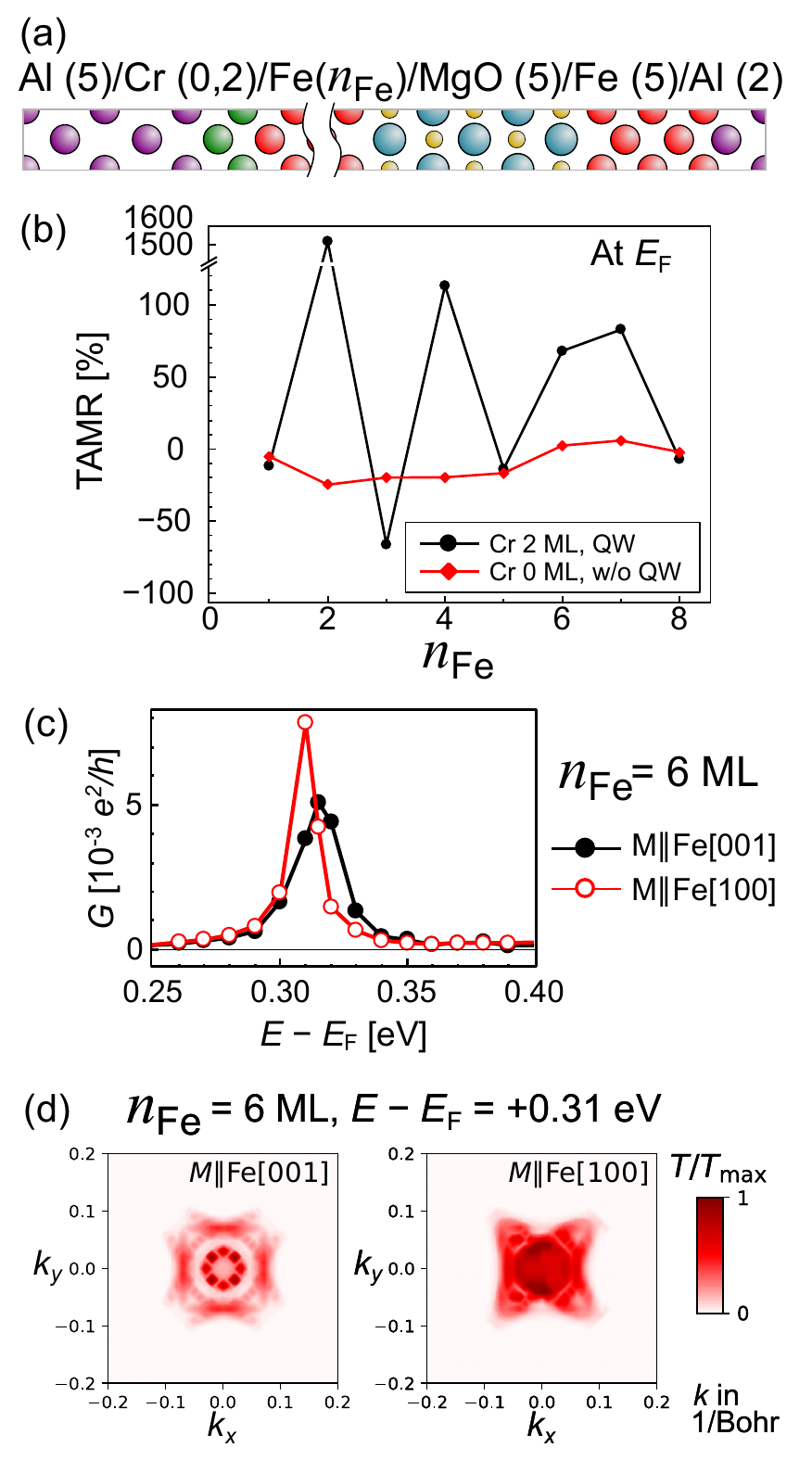}
    	\caption{Density-functional theory calculations of QW-TAMR. (a) A depiction of the calculated structure. (b) The thickness dependence shows that TAMR oscillates with $n_\mathrm{Fe}$ for the QW case. (c) The energy dependence of average transmittance in Cr(2 ML)/Fe(6 ML)/MgO. The QWRS level has a shift induced by the rotation of $M$. (d) The calculated $T$--$k_\parallel$ map at resonance condition ($E_F + 0.31$ eV). The map is calculated for $M$ oriented along the out-of-plane (left panel) and inplane (right panel) directions.
    	}
    	\label{fig:calc_res}
    \end{figure}
    
     To further elucidate the main origin of QW-TAMR, we calculated the spin-dependent transport properties in the stack of Cr/Fe/MgO/Fe\hkl(001) by first-principles calculations, using the \textsc{Quantum ESPRESSO} package \cite{giannozzi_2009,giannozzi_2017}. Figure \ref{fig:calc_res}(a) depicts the simulated structure of Al (5)/Cr (0 or 2)/Fe (1--8)/MgO (5)/Fe (5)/Al (2) (thickness in MLs), where the number of Fe MLs ($n_\text{Fe}$) was varied from 1 to 8.
     The density-functional theory calculations are conducted with fully-relativistic ultrasoft pseudopotentials, and under the local-density approximation \cite{choi_1999,smogunov_2004}. The number of $k$ points was taken to be $20\times 20 \times 1$ for electronic structure calculations, and $100 \times 100$ for tunneling conductance calculations. Methfessel-Paxton smearing with a broadening parameter of \SI{0.01}{Ry} was used. The cutoff energy for the wavefunction was set to \SI{30}{Ry}, and for the charge density to \SI{300}{Ry}.
    The atomic positions are fully optimized in the calculations. The transmittance $T$ is calculated with $M$ oriented either along the Fe \hkl[001] or Fe \hkl[100] directions. The aluminium layers are included as an unpolarized bath of $\Delta_1$ electrons. We confirmed the formation of the QWRSs by confinement between the 2-ML Cr layer and the MgO barrier, while allowing for a non-vanishing conductance. The choice of MgO in the calculations is representative of the main effects in our QW. The tunneling transport is mainly through the oxygen sublattice \cite{zhang_2003}, and the cation-disordered MgAl$_2$O$_4$ is equivalent to a lattice-matched MgO and the band-folding effects can be ignored \cite{sukegawa_2012,miura_2012}.
     
     Figure \ref{fig:calc_res}(b) shows the dependence of the calculated TAMR on $n_\text{Fe}$, at the Fermi level. For the case without a Cr insertion (w/o QW), the TAMR is relatively small and shows a smooth dependence on $n_\text{Fe}$. On the other hand, the formation of QWRS by Cr insertion, QW-TAMR has a much larger magnitude, and oscillates with $n_\text{Fe}$, similar to the experiments. We find the origin of the QW-TAMR to be the anistropic shift in QWRS. The interfacial magnetocrystalline anisotropy of Fe/MgO interface changes the potential well leading to a shift in QWRS, and therefore QW-TAMR is observed.
     Figure \ref{fig:calc_res}(c) shows the energy dependence of the total conductance at $n_\mathrm{Fe} =$ 6 ML. We find that the QWRSs are mainly formed by the Fe($d_{3z^2-r^2}$) states with the $\Delta_1$ symmetry at the Fe/oxide and Cr/Fe interfaces. The $\Delta_1$ QW states are confined by the potential of the interfacial bonding between Fe $d_{3z^2-r^2}$ and O $p_z$ orbitals at one side, and Fermi surface mismatch between Fe and Cr on the other side. The resonant tunneling through QWRS is observed at $E = 0.315$ eV, where $E = 0$ eV corresponds to the Fermi level. The rotation of $M$ causes a shift in the energy of QWRS. The transmittance peak is shifted and increases in magnitude at $M\!\!\parallel\!\!\mathrm{Fe} \hkl[100]$ [Fig.~\ref{fig:calc_res}(c)], which is consistent with the experiments. We note that the calculated QWRS energy positions do not match quantitatively to the experimental positions. This difference can be attributed to the different atomic configuration between the experiments and the calculations, especially at the Cr layer.
     
     We analyze the components of conduction anisotropy in Table \ref{tab:G}, which lists the spin-conserved ($G_{\uparrow\uparrow}$, $G_{\downarrow\downarrow}$) and spin-flip ($G_{\uparrow\downarrow}$, $G_{\downarrow\uparrow}$) conductance components at $k_\parallel = (0,0)$. The conduction is dominated by the resonant majority-to-majority spin tunneling in the present QW structure, and the spin-flip channel shows a very small magnitude. It is the modulation of QWRS that causes the large anisotropy in the spin-conserved tunneling channel $G_{\uparrow\uparrow}$, hence the large QW-TAMR effect. On the other hand, in the case of thick non-QW Fe electrodes, the spin-flip components through IRSs were the origin to the small non-QW TAMR effect \cite{chantis2007,khan_2008,lu_2012}.
     
     \begin{table}[!ht] 
    \caption{\label{tab:G}The spin-resolved conductance components of the QW calculated at $k_\parallel = (0,0)$, $E = 0.315$ eV, where Fe thickness is 6 MLs. All the quantities are in units of conductance quantum $e^2/h$, where $e$ and $h$ are the electron charge and Planck's constant, respectively. }
    
        \centering
        
        \begin{tabular}{|l|c|c|c|c|}
            \hline
             & $G_{\uparrow\uparrow}$ & $G_{\uparrow\downarrow}$ & $G_{\downarrow\uparrow}$ & $G_{\downarrow\downarrow}$ \\
            \hline
            $M\!\!\parallel\!\!\mathrm{Fe} \hkl[001]$ & \num{0.6747} & \num{1.10e-7} & \num{1.10e-7} & \num{7.63e-11} \\
            \hline
            $M\!\!\parallel\!\!\mathrm{Fe} \hkl[100]$ & \num{0.4443} & \num{1.04e-4} & \num{1.04e-4} & \num{2.43e-8} \\
            \hline
        \end{tabular}
    \end{table}
    
    The effect of $M$ rotation on QWRS can be understood further by looking into the $k_\parallel$-resolved transmittance map [Fig.~\ref{fig:calc_res}(d)]. For $M\!\!\parallel\!\!\mathrm{Fe}\hkl[001]$, the $T$--$k_\parallel$ map is fourfold symmetric, with narrow resonant conduction peaks. When $M$ lies along $\mathrm{Fe}\hkl[100]$ direction, SOC causes a large deformation and broadening of the $T$--$k_\parallel$ map, and the rotational symmetry is broken. Mainly, the QW resonant transmittance at the $\Gamma$ point is strongly modulated, while the IRSs at $k_\parallel \neq 0$ are also shifted, due to the change of the QW potential of Cr/Fe/MgO. In this case, $T$--$k_\parallel$ map of $M\!\!\parallel\!\!\mathrm{Fe}\hkl[100]$ state loses the mirror-symmetry around the $k_y$ \hkl[010] axis. If the Rashba SOC is the origin of this QW-TAMR, the bands with $k \neq 0$ should divert along the $k$ direction orthogonal to the magnetization \cite{chantis2007,khan_2008}, and therefore break the mirror-symmetry around the $k_x$ \hkl[100] axis. In our case, the asymmetry around $k_x$ axis in the $T$--$k_\parallel$ map is very small, indicating that the Rashba SOC contribution to the QW-TAMR is negligible. 
    We note that the band-theory origin of bulk Fe anisotropic magnetoresistance is related to the spin-flip scattering between $\Delta_2$-symmetry states \cite{khan_2008,zeng_2020}. Hence, a ballistic bulk-Fe AMR does not contribute to TAMR in MgO MTJs \cite{khan_2008}
    
In summary, we observed a large QW-TAMR effect, where the rotation of QW magnetization modulates the majority-to-majority resonant transmission. We measured QW-TAMR in lattice-matched Cr/Fe/MgAl$_2$O$_4$ resonant tunnel junctions, in which the resonant states are symmetry-selected for $\Delta_{1,\uparrow}$ electronic states. This QW-TAMR effect is present up to high temperatures, $>\SI{380}{\kelvin}$.
 Analogous to semiconductor QW transistors, the magnetization angle has the same role as a transistor's gate voltage, where both can control the energy positions of quantization levels. We suggest that the magnetic gating of QWs by the spin-degree-of-freedom can be applied to new electronic and optoelectronic devices, such spin-polarized light sources and detectors \cite{yu_2016}. 
 \paragraph*{Acknowledgments}
This work was partly supported by the ImPACT Program of the Council for Science, Technology and Innovation (Cabinet Office, Government of Japan), and the JSPS KAKENHI Grant Number JP16H06332.

\paragraph*{Additional Information}
Supplementary Information is available for this paper.

\paragraph*{Authors contributions}
M.A. is the corresponding author and conceived the project.  M.A., Q.X., and M.B. conducted the experiments, and Y.M. conducted the first-principles calculations. M.A., Q.X., Y.M., M.B., K.M., S.K., H.S., and S.M. elaborated on the results, and formulated the discussions. M.A. wrote the manuscript with input from Y.M. S.M., H.S., and the other authors.

\paragraph*{Competing interests}
The authors declare no competing financial interests.

\paragraph*{Data and materials availability}
The presented data and related supporting data may be requested from the authors, upon a reasonable request.

\newcommand{\noopsort}[1]{}
%

\newpage
\clearpage
\onecolumngrid
\begin{center}
\textbf{\large Supplementary Materials: Quantum-well tunneling anisotropic magnetoresistance above room temperature}
\end{center}
\setcounter{section}{0}
\setcounter{equation}{0}
\setcounter{figure}{0}
\setcounter{table}{0}
\setcounter{page}{1}
\makeatletter
\renewcommand{\thesection}{S\arabic{section}}
\renewcommand{\theequation}{S\arabic{equation}}
\renewcommand{\thefigure}{S\arabic{figure}}
\renewcommand{\bibnumfmt}[1]{[S#1]}
\renewcommand{\citenumfont}[1]{S#1}

\section{Origin of the fourfold component in TAMR-$\psi$ curves}\label{sec:sup_a4}
In this section, we show that the fourfold component of TAMR-$\psi$ curves observed in Figs.~1(b), 2(b)
of the main text is due to the misalignment between $H$ and $M$ during rotation, under the influence of magnetic anisotropy. This is similar to the apparent fourfold component in torque magnetometry, when $H$ is not much larger than the anisotropy field $H_K$ \cite{chikazumi_2009_4fold_s}. This is more prominent in the present case. The perpendicular magnetic anisotropy is very large at Fe/oxide epitaxial interface, with an effective anisotropy field ($H_K = 2 K_\mathrm{eff} / M_s$) reaching 20 kOe at room temperature \cite{lambert_2013_s,koo_2013_s,xiang_2017_s}. At the measurement temperature of 5 K, $H_K$ is more than 50 kOe [Fig.~2(a)]. 
A non-practical magnetic field of 400--500 kOe would be needed to remove this extrinsic fourfold component. The accurate estimation of $K_\mathrm{eff}$ in our study is complicated by the presence of QW-TAMR, hindering the accurate extraction of $M$-$H$ curves from $G$-$H$ measurements. Without the knowledge of $K_\mathrm{eff}$, it is difficult to find QW-TAMR's intrinsic twofold and fourfold components utilizing the procedures of Refs.~\cite{gao2007_s,tao_2018_s}.

Using an alternative procedure, we could deduce the intrinsic twofold part related to the rotation of $M$ from the measurements on twofold and fourfold components obtained from the rotation of $H$. The following analysis is based on assuming that QW-TAMR has solely a twofold symmetry in $\theta$. Following the same procedure as Ref.~\cite{chikazumi_2009_4fold_s}, we define the direction cosines of $M$ and $H$ projected on the $z$-axis as $\alpha_3$ and $\beta_3$, respectively. We define TAMR$(\theta)$ as:

\begin{align}
\mathrm{TAMR}(\theta) &= \frac{G(\theta)}{G(90^\circ)} - 1 \nonumber \\
					&= 2 A_{2\theta} \cos^2 \theta \nonumber \\
                    &\equiv 2 A_{2\theta} \alpha_3^2,
\end{align}
where $\alpha_3$ can be related to $\beta_3$ by the following equality:

\begin{equation}
\alpha_3 = (1+p+p^2) \beta_3 - (p+ \frac{7}{2} p^2) \beta_3^3 + \frac{5}{2} p^2 \beta_3^5 + \cdots,
\label{eq:alpha_beta}
\end{equation}
and:

\begin{equation}
p = \frac{2 K_\mathrm{eff}}{M_s H} = \frac{H_K}{H} \ll 1.
\end{equation}

After retaining the linear terms in $p$, $\alpha_3^2$ can be approximated to:
\begin{align}
\alpha_3^2 & \approx (1+2p+p^2) \beta_3^2 - 2(p+p^2) \beta_3^4 + p^2 \beta_3^6 \nonumber \\
		   & \approx (1+2p) \beta_3^2 - 2p \beta_3^4.
\end{align}

By trigonometric identities, we can change TAMR$(\theta)$ into TAMR$(\psi)$ in terms of twofold and fourfold components as follows:
\begin{align}
\mathrm{TAMR}(\psi) & \approx A_{2\theta} \left[ (1-p) + (1-2p) \cos 2\psi - p \cos 4\psi \right] \nonumber \\
					& \equiv \mathrm{const.} + a_{2\psi} \cos 2\psi + a_{4\psi} \cos 4\psi,
\end{align}
where:
\begin{equation}
a_{2\psi} \approx A_{2\theta} (1-2p), \quad \mathrm{and} \quad a_{4\psi} \approx A_{2\theta} (-p). \label{eq:a2a4}
\end{equation}

A field-dependent fourfold component in $\psi$ appeared, even though we assumed only a twofold dependence of QW-TAMR on $\theta$. The interinsic $A_{2\theta}$ can be found as:
\begin{equation}
A_{2\theta} \approx a_{2\psi} - 2 a_{4\psi}. \label{eq:A2}
\end{equation}

\begin{figure}
	\includegraphics[width=0.7\textwidth]{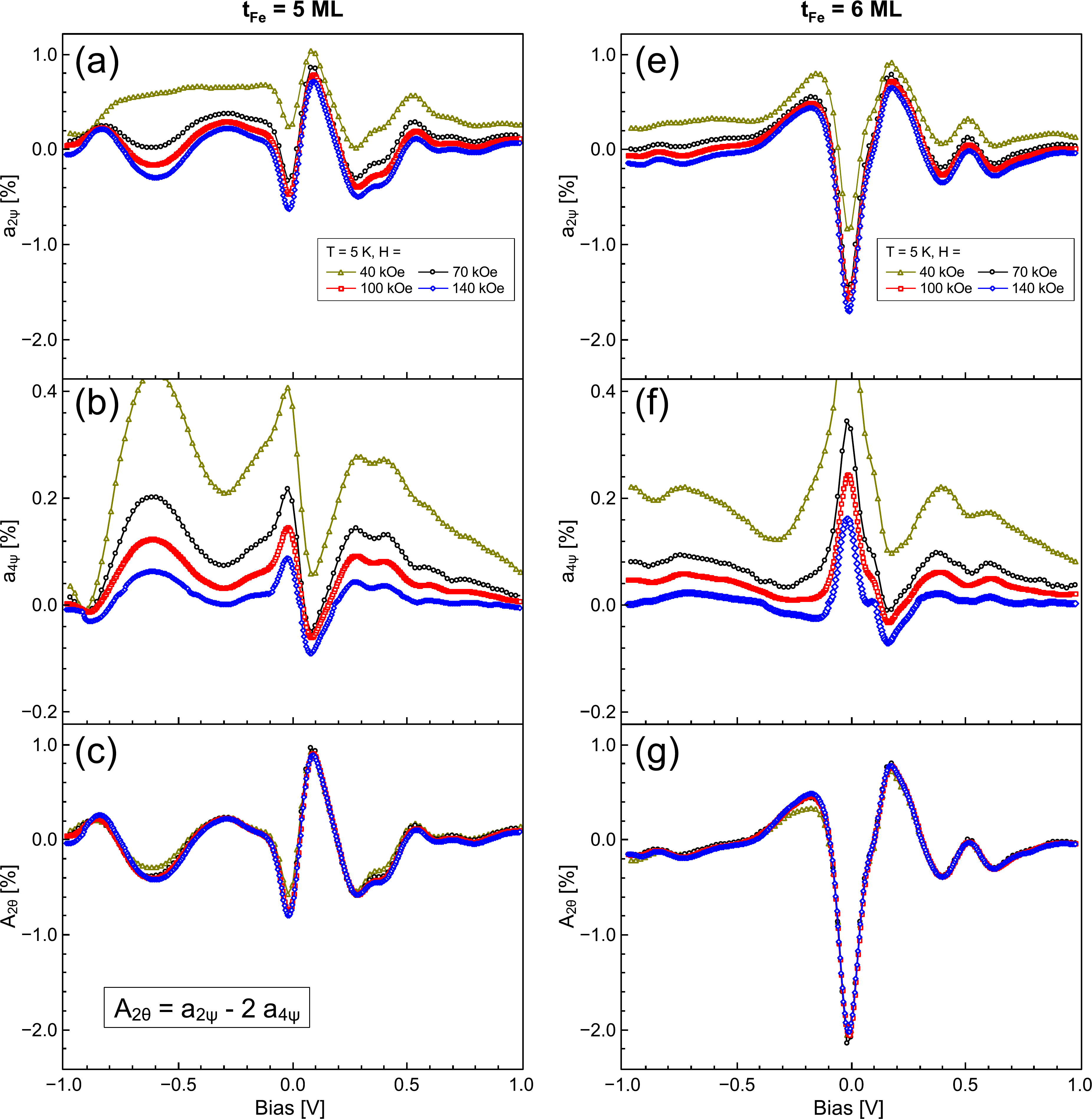}
	\caption{Bias dependence of (a,e) $a_{2\psi}$, (b,f) $a_{4\psi}$, and (c,g) $A_{2\theta}$ under various magnetic fields. The thickness of Fe layer is (a,b,c) $t_\mathrm{Fe}$ = 5 ML, and (e,f,g) $t_\mathrm{Fe}$ = 6 ML.}
	\label{fig:a2a4}
\end{figure}

To test the validity of this approximation, we measured the bias dependence of $a_{2\psi}$ and $a_{4\psi}$ at $H$ = 40, 70, 100, 140 kOe, in the QW samples of the main text [Fig.~\ref{fig:a2a4}]. In accordance with the conclusions from Eqs.~\ref{eq:a2a4}, $a_{4\psi}$ has an inverted character to that of $a_{2\psi}$, and also diminishes in magnitude as $H$ increases. However, even at $H$ = 140 kOe, $a_{4\psi}$ has a non-zero magnitude resembling $a_{2\psi}$ character. Therefore, it seems that an intrinsic fourfold component is negligible in our QW system. Figs.~\ref{fig:a2a4}(c,g) show that the deduced $A_{2\theta}$ has a good agreement between various $H$ magnitudes. The approximation of Eq.~\ref{eq:A2} is still valid even at 40 kOe, which is close to $H_K$. We conclude by stating that Eq.~\ref{eq:A2} allows the estimation of $A_{2\theta}$, even without knowing the magnitude of magnetic anisotropy energy. Finally, we emphasize that the field dependence of the fourfold component should always be checked before inferring on the existence of high-order anisotropies.

\section{QW-TAMR in the out-of-plane and inplane rotation planes}\label{sec:sup_plane}

\begin{figure}
    \includegraphics[width=0.7\textwidth]{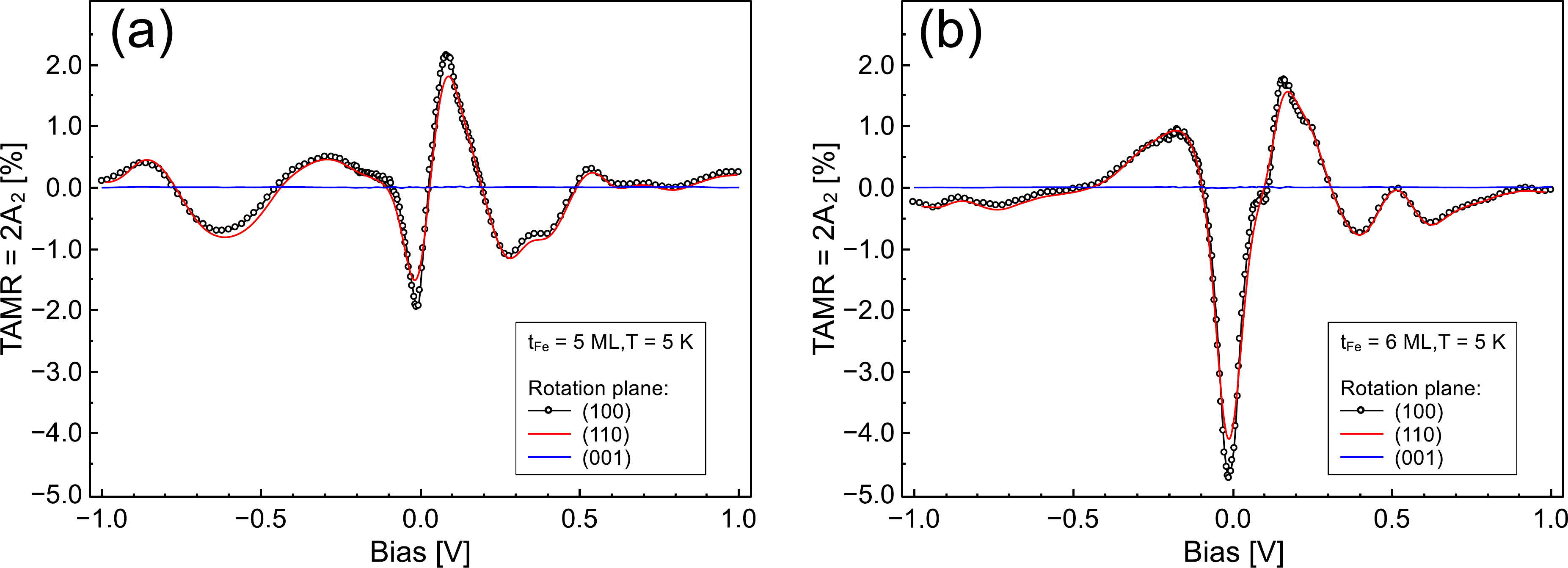}
	\caption{Bias dependence of QW-TAMR in different rotation planes. (a) $t_\mathrm{Fe}$ = 5 ML, and (b) $t_\mathrm{Fe}$ = 6 ML. In the \hkl(001) inplane rotation, QW-TAMR is negligible.}
	\label{fig:a2_plane}
\end{figure}

We measured the effect of rotation plane on QW-TAMR spectra [Fig.~\ref{fig:a2_plane}]. We found that the QW-TAMR spectra are the same the for $M$ rotation in Fe \hkl(100) and \hkl(110) planes. Moreover, the inplane \hkl(001) rotation produces zero TAMR effect. In bulk bcc Fe, \hkl(100) and \hkl(110) planes are not equivalent, therefore if the QWRS shift was of bulk origin, a difference should have been found. On the other hand, the interfacial asymmetry, and hence the crystal field, is much larger at the Fe-O bond. Therefore, QW-TAMR was observed only in the out-of-plane rotation direction.

\section{Oscillation of TAMR with QW thickness}\label{sec:sup_osc}

 As the QW thickness increases, the QWRS of the next nodal-mode ($m+1$) decreases in energy until it passes through the Fermi level. The thickness period of such crossing will be commensurate with half the Fermi wavelength ($\lambda_\mathrm{F}/2$). This is also consistent with simplified phase-accumulation models, since the phase shifts at QW boundary reflections do not change significantly with QW thickness \cite{*[{}] [{; and the references therein.}] smith_1994_s}. The Fe $\Delta_{1,\uparrow}$ states cross the Fermi level in the middle ($k \approx 0.47$) of $\Gamma$-$H$ line. 
Therefore, the period of QWRS crossings at the Fermi level will be $2.1$ ML, \emph{i.e.} commensurate with the parity of Fe MLs number.

\begin{figure}[ht!]
 \vspace{1\baselineskip}
	\includegraphics[width=0.7\textwidth]{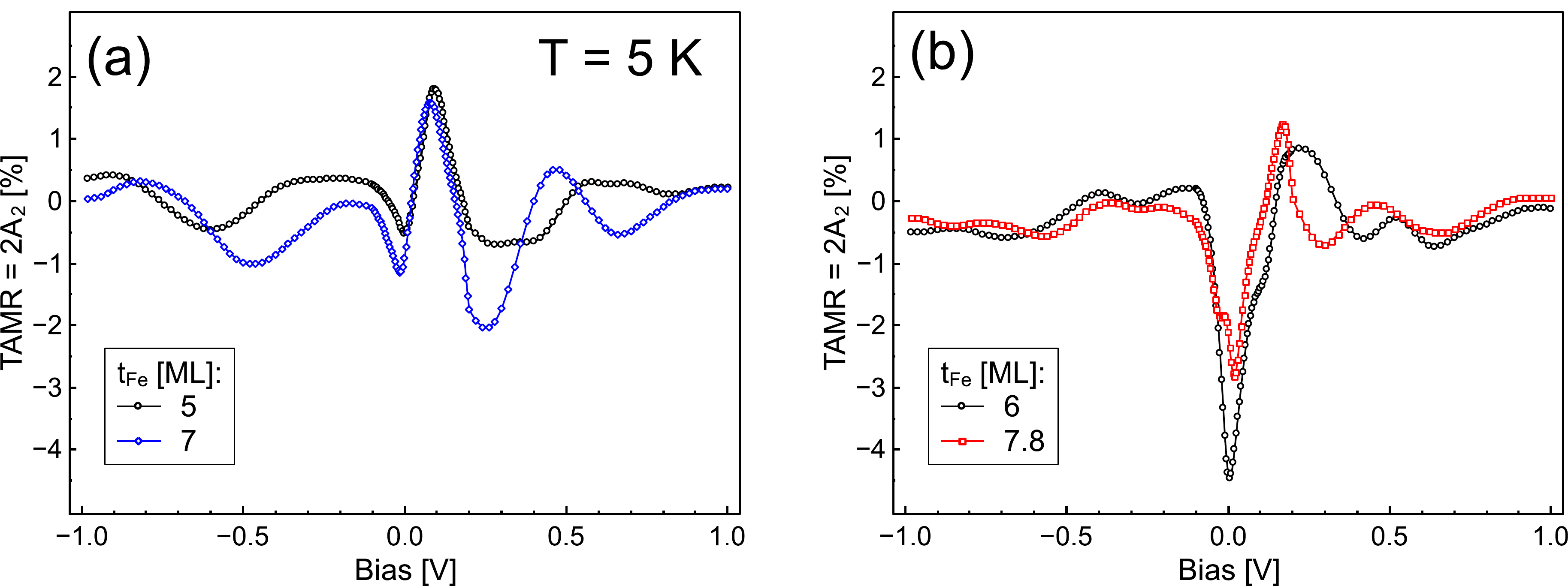}
	\caption{The oscillation of TAMR in Fe-QW thickness. (a) Odd number of Fe MLs (b) Even number of Fe MLs.}
	\label{fig:tFeOsc}
\end{figure}

We measured the effect of the parity of number of QW-Fe MLs on QW-TAMR. In another sample different from the one of the main text, $t_\mathrm{Fe}$ was linearly varied in a wedge-shaped film stack MgO \hkl(001) substrate/MgO (5)/Cr (30)/Fe ($t_\mathrm{Fe}$)/MgAl$_2$O$_4$ (2)/Fe (10)/Ru (15) (thicknesses in nm). The film was microfabricated into junctions, and from the peaks of $dI/dV$--$V$ spectra, the junctions with exact integer number of MLs were confirmed. The thicknesses $t_\mathrm{Fe} = 5, 6, 7, 7.8$ ML were chosen for QW-TAMR measurements. QW-TAMR spectra at 5 K show the same character for each parity of $t_\mathrm{Fe}$, \emph{i.e.}~each pair of (5, 7) ML and (6, 7.8) ML have similar QW-TAMR spectra [Fig.~\ref{fig:tFeOsc}]. Therefore, QW-TAMR has an oscillatory behavior with $t_\mathrm{Fe}$, due to formation of QW levels near or far from Fermi level.


%

\end{document}